\documentclass[twoside]{article}
\usepackage{fleqn,espcrc2}
\usepackage[dvips]{graphicx}
\usepackage[figuresright]{rotating}
\usepackage{epsf}
\usepackage{amssymb}
\def\bc{\begin{center}}
\def\ec{\end{center}}
\def\beq{\begin{equation}}
\def\eeq{\end{equation}}
\def\bs{\begin{slide}}
\def\es{\end{slide}}
\newcommand{\bmath}{\begin{displaymath}}
\newcommand{\emath}{\end{displaymath}}
\newcommand{\beqn}{\begin{eqnarray}}
\newcommand{\eeqn}{\end{eqnarray}}
\newcommand{\beqns}{\begin{eqnarray*}}
\newcommand{\eeqns}{\end{eqnarray*}}
\newcommand{\ba}{\begin{array}{c}} 
\newcommand{\bat}{\begin{array}{cc}} 
\newcommand{\ea}{\end{array}}

\newcommand{\lsim}{\stackrel{<}{_\sim}}

\newcommand{\Frac}[2]{\frac{\displaystyle #1}{\displaystyle #2}}
\newcommand{\tppp}{\tau\to\pi\pi\pi\nu_\tau}
\newcommand{\tpppc}{\tau^-\to\pi^+\pi^-\pi^-\nu_\tau}
\newcommand{\tpppn}{\tau^-\to\pi^-\pi^0\pi^0\nu_\tau}

\title{$\tau^- \rightarrow (\pi \pi \pi)^- \nu_{\tau}$~: Theory
versus Experiment
\thanks{IFIC/03$-$58 report. Talk given by J. Portol\'es at
the High--Energy Physics International
Conference on Quantum Chromodynamics, 2--9 July (2003), Montpellier (France).}
}

\author{D. G\'omez Dumm \address[Argentina]{IFLP, Depto. de F\'{\i}sica, 
Universidad Nacional de la Plata, \\ C.C. 67, 1900 La Plata, Argentina},
        A. Pich \address[Espanya]{Departament de F\'{\i}sica Te\`orica, IFIC,
CSIC-Universitat de Val\`encia, \\ Apt. Correus 22085, E-46071 Val\`encia,
Spain}, 
        J. Portol\'es \addressmark[Espanya]} 
       
\begin{document}

\begin{abstract}
We analyse $\tppp$ decays within the framework of the resonance
chiral theory of QCD. We have worked out the relevant Lagrangian that describes
the axial--vector current hadronization contributing to these processes,
and the new coupling constants that arise have been constrained by
imposing the asymptotic behaviour of the corresponding spectral function
within QCD. Hence we compare the theoretical framework with the experimental
data, obtaining a good quality fit from the ALEPH spectral function and
branching ratio. We also get values for the mass and on--shell width
of the $a_1$(1260) resonance, and provide the $\tppp$ structure functions
that have been measured by OPAL and CLEO-II finding an excellent agreement.
\vspace{1pc}
\end{abstract}

\maketitle

\section{Introduction}

The hadronization of currents in exclusive processes provide a
detailed knowledge on the strong interacting mechanisms driven by
non--perturbative QCD. Within this framework $\tau$ decays into hadrons
allow to study the properties of the vector and axial--vector QCD 
currents, and yield relevant information on the dynamics of the
resonances entering into the processes. At very low energies, 
typically $E \ll M_{\rho}$ (being $M_{\rho}$ the mass of the 
$\rho$(770)), chiral perturbation theory ($\chi$PT) \cite{bi:1} 
is the corresponding effective theory of QCD. However the decays
$\tppp$, through their full energy spectrum, happen to be driven by 
the $\rho$(770) and $a_1$(1260) resonances, mainly, in an energy region 
where the invariant hadron momentum approaches the masses of the
resonances ($\sqrt{Q^2} \sim M_{\rho}$). Consequently $\chi$PT is no 
longer applicable to the study of the whole spectrum but only to the
very low energy domain \cite{bi:3}. Until now the standard way of 
dealing with these decays has been to use ${\cal O}(p^2)$ $\chi$PT
to fix the normalization of the amplitudes in the low energy region and,
accordingly, to include the effects of vector and axial--vector meson 
resonances by modulating the amplitudes with {\rm ad hoc} Breit--Wigner
functions \cite{bi:4,bi:5}. However we have seen \cite{nos,Victoria}
that this modelization
is not consistent with ${\cal O}(p^4)$ $\chi$PT, a fact that could
spoil any outcome provided by the analysis of experimental data using
this procedure.
\par
Lately several experiments have collected good quality data on 
$\tppp$, such as branching ratios and spectra \cite{bi:6} or 
structure functions \cite{bi:8}. Their analysis within a 
model--independent framework is highly desirable if one wishes to 
collect information on the hadronization of the relevant QCD currents.
At energies $E \sim M_{\rho}$ the resonance mesons are active degrees
of freedom that cannot be integrated out, as in $\chi$PT, and they have
to be properly included into the relevant Lagrangian. The
procedure is ruled by the approximated chiral symmetry of QCD under
$SU(3)_{\mathrm L} \otimes SU(3)_{\mathrm R}$, that drives the 
interaction of Goldstone bosons (the lightest octet of pseudoscalar
mesons), and the $SU(3)_{\mathrm V}$ assignments of the resonance
multiplets. Its systematic arrangement has been put forward in 
Refs.~\cite{bi:10,bi:11} as the Resonance Chiral Theory (R$\chi$T).
A complementary tool is the large number of colours ($N_C$) limit of QCD.
It has been pointed out \cite{bi:12} that the inverse of the number
of colours of the gauge group $SU(N_C)$ could provide the expansion
parameter involved in the perturbative treatment of the amplitudes.
Indeed large--$N_C$ QCD shows features that resemble, both qualitatively
and quantitatively, the $N_C=3$ case.
\par
In Ref.~\cite{nos} we have performed an analysis of the $\tppp$ decays
using de above--mentioned tools, and we present in the following
its main results.
\vspace*{-0.314cm}
\section{The Resonance Effective Theory of QCD}

The final hadron system in the $\tppp$ decays spans a wide energy
region $3 m_{\pi}  \lsim  E  \lsim  M_{\tau}$ that is heavily
populated by resonances. As a consequence an effective theory 
description of the full energy spectrum requires to include the
resonances as active degrees of freedom. R$\chi$T is the appropriate
framework to work with and, accordingly, we consider the Lagrangian
\begin{eqnarray}
\label{eq:ret}
{\cal L}_{\rm R\chi T} \! \!   & = \! \! \!  &  
\frac{F^2}{4}\langle u_{\mu}
u^{\mu} + \chi _+ \rangle \, + \, \frac{F_V}{2\sqrt{2}} \langle V_{\mu\nu}
f_+^{\mu\nu}\rangle \nonumber \\
& & + \, i \,\Frac{G_V}{\sqrt{2}} \langle V_{\mu\nu} u^\mu
u^\nu\rangle  \, + \, 
\frac{F_A}{2\sqrt{2}} \langle A_{\mu\nu}
f_-^{\mu\nu}\rangle \,\nonumber \\
& &  + \, {\cal L}_{\rm kin}^{\rm V} \, + \,  {\cal L}_{\rm kin}^{\rm A} \, + 
\, \sum_{i=1}^{5}  \, \lambda_i  \,
{\cal O}^i_{\rm VAP} \, ,
\end{eqnarray}
with $\lambda_i$ unknown real adimensional couplings, and
the operators ${\cal O}^i_{\rm VAP}$ are given by
\begin{eqnarray}
\label{lag2}
{\cal O}^1_{\rm VAP} &  = & \langle \,  [ \, V^{\mu\nu} \, , \, 
A_{\mu\nu} \, ] \,  \chi_- \, \rangle \; \; , \nonumber \\
{\cal O}^2_{\rm VAP} & = & i\,\langle \, [ \, V^{\mu\nu} \, , \, 
A_{\nu\alpha} \, ] \, h_\mu^{\;\alpha} \, \rangle \; \; , \\
{\cal O}^3_{\rm VAP} & = &  i \,\langle \, [ \, \nabla^\mu V_{\mu\nu} \, , \, 
A^{\nu\alpha}\, ] \, u_\alpha \, \rangle \; \; ,  \nonumber \\
{\cal O}^4_{\rm VAP} & = & i\,\langle \, [ \, \nabla^\alpha V_{\mu\nu} \, , \, 
A_\alpha^{\;\nu} \, ] \,  u^\mu \, \rangle \; \; , \nonumber \\
{\cal O}^5_{\rm VAP} & =  & i \,\langle \, [ \, \nabla^\alpha V_{\mu\nu} \, , \, 
A^{\mu\nu} \, ] \, u_\alpha \, \rangle \nonumber \; \; ,
\end{eqnarray}
where $h_{\mu \nu} = \nabla_{\mu} u_{\nu} + \nabla_{\nu} u_{\mu}$ and
the notation is that of Ref.~\cite{bi:10}. Notice that we use the antisymmetric
tensor formulation to describe the spin 1 resonances and, consequently,
we do not consider the ${\cal O}(p^4)$ chiral Lagrangian of pseudoscalars
\cite{bi:11}.

\section{Axial--vector current form factors in $\tppp$}

The decay amplitude for the $\tpppc$ and $\tpppn$ processes can be
written as 
\begin{equation} {\cal M}_{\pm} \; = \; - \,
\frac{G_F}{\sqrt{2}} \, V_{ud} \, \bar u_{\nu_\tau}
\gamma^\mu\,(1-\gamma_5) u_\tau\, T_{\pm \mu} \; \; ,
\end{equation}
where $T_{\pm \mu}$ is the hadronic matrix
element of the participating $V_{\mu} - A_{\mu}$ QCD currents. In the isospin
limit there is no contribution of the vector current to these processes and,
therefore, only the axial--vector current $A_\mu$ appears~:
\begin{equation}
T_{\pm \mu}(p_1,p_2,p_3)  = 
 \langle  \pi_1(p_1)\pi_2(p_2)\pi^{\pm}(p_3)  |   A_\mu  |  
 0  \rangle \, ,
\label{matelem}
\end{equation}
being $\pi^+$ the one in $\tpppc$ and  $\pi^-$ that in $\tpppn$. The hadronic
tensor can be written in terms of three form factors, $F_1$, $F_2$ and
$F_P$, as \cite{bi:14}
\begin{equation}
T^{\mu} \; = \; V_1^\mu\,F_1 \, + \,  V_2^\mu\,F_2 \, + \,  
V_P^\mu\,F_P \; \; ,
\label{tmu}
\end{equation}
where
\begin{eqnarray}
V_1^\mu & = & \left( \, g^{\mu \nu} \, - \, 
\Frac{Q^{\mu} Q^{\nu}}{Q^2} \, \right) \, ( \, p_1 - p_3 \, )_{\nu} \; \; ,
\nonumber \\
V_2^\mu & = & \left( \, g^{\mu \nu} \, - \, 
\Frac{Q^{\mu} Q^{\nu}}{Q^2} \, \right) \, ( \, p_2 - p_3 \, )_{\nu} \; \; ,
\nonumber \\
V_P^\mu & = & Q^\mu \, = \,  p_1^\mu + p_2^\mu + p_3^\mu\; \; .
\end{eqnarray}
The form factors $F_1$ and 
$F_2$ have a transverse structure in the 
total hadron momenta $Q_{\mu}$ and drive a
$J^P=1^+$ transition. Bose symmetry under interchange of the
two identical pions in the final state demands that 
$F_1(Q^2,s,t) = F_2(Q^2,t,s)$ where $s=(p_1+p_3)^2$ and $t=(p_2+p_3)^2$.
Meanwhile $F_P$ accounts for a $J^P=0^-$
transition that carries pseudoscalar degrees of 
freedom and vanishes with the square of the pion mass. Hence its
contribution to the decay processes will be very much suppressed and we 
will not consider it in the following. 
\par
\begin{figure}[!htb]
\hspace*{0.7cm} 
\includegraphics*[scale=0.8,clip]{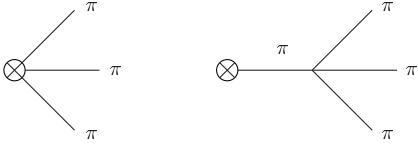}
\vspace*{-0.7cm}
\caption{\label{fig:f1}Diagrams contributing to the hadronic
amplitude  $T_{\pm \mu}$ in ${\cal O}(p^2)$ $\chi$PT. }
\vspace*{-0.5cm}
\end{figure}
In the low $Q^2$ region, the matrix element in Eq.\ (\ref{matelem}) can be
calculated using $\chi$PT. At ${\cal O}(p^2)$ one has two contributions, 
arising from the diagrams in Fig.\ \ref{fig:f1}. 
The sum of both graphs yields
\begin{equation}
T_{\pm\mu}^{\chi} \, = \, 
 \mp \frac{2\sqrt{2}}{3 F}\left\{ V_{1\mu} + V_{2\mu} \right\}  \; \; .
\label{lowq}
\end{equation}
\begin{figure}[!htb]
\includegraphics*[scale=0.65,clip]{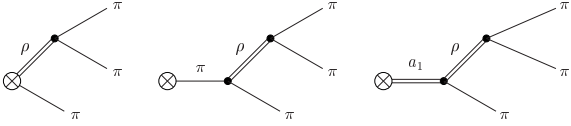} 
\vspace*{-0.7cm}
\caption{\label{fig:f2} Resonance--mediated diagrams contributing
to $T_{\pm \mu}$.}
\vspace*{-0.5cm}
\end{figure}
We now include the resonance--mediated contributions
to the amplitude, to be evaluated through the interacting terms
in ${\cal L}_{\rm R\chi T}$ Eq.~(\ref{eq:ret}). 
The relevant diagrams to be taken into account are those shown in 
Fig.\ \ref{fig:f2}. We get 
\begin{eqnarray}
\label{eq:t1r}
T_{\pm\mu}^{R} & = & \mp \frac{\sqrt{2}\,F_V\,G_V}{3\,F^3} \times 
\\ & &  
\left[ \alpha(Q^2,s,t) \, V_{1\mu} + \alpha(Q^2,t,s) \, V_{2\mu} \right]
 \nonumber \\
 & & \pm \Frac{4 \, F_A \, G_V}{3 \,F^3} \, 
 \Frac{Q^2}{Q^2-M_A^2} \, \times \nonumber 
 \\
 & & \left[ \beta(Q^2,s,t) \, V_{1 \mu} + \beta(Q^2,t,s) \, V_{2 \mu}
 \right] \; . \nonumber
\end{eqnarray}
The functions $\alpha(Q^2,s,t)$
and $\beta(Q^2,s,t)$ are given explicitly in Ref.~\cite{nos} and depend
on three combinations of the $\lambda_i$ couplings in 
${\cal L}_{\rm R \chi T}$
in Eq.~(\ref{eq:ret}), that we call $\lambda_0$, $\lambda'$ and 
$\lambda''$.
\par
The form factors in Eq.\ (\ref{eq:t1r}) include 
zero--width $\rho$(770) and
$a_1$(1260) propagator poles, leading to divergent phase--space integrals 
in the calculation of the $\tppp$ decay width as the kinematical variables
go along the full energy spectrum. The result can be regularized
through the inclusion of resonance widths, which means to go beyond the
leading order in the $1/N_C$ expansion, and implies the introduction of
some additional theoretical inputs. This issue has been analysed in detail 
within the resonance chiral effective theory in Ref.~\cite{bi:15} and, 
accordingly, we include off--shell widths for both resonances \cite{nos}.
\begin{figure}[!htb]
\hspace*{-0.5cm}
\includegraphics*[angle=-90,scale=0.33,clip]{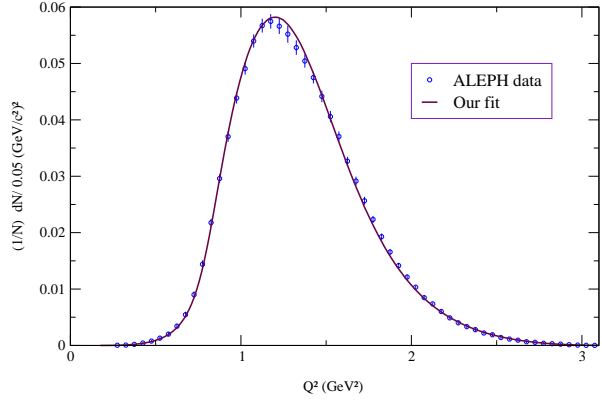}
\vspace*{-1cm}
\caption{\label{fig:f3} Fit to the ALEPH data for the normalized
$\tau^- \rightarrow \pi^+ \pi^- \pi^- \nu_{\tau}$. }
\vspace*{-0.5cm}
\end{figure}

\section{QCD constraints on the coupling constants}

A look to our results above shows that we have six unknown couplings 
(or combination of couplings), namely~: $F_V$, $G_V$, $F_A$, $\lambda_0$,
$\lambda'$ and $\lambda''$. 
The QCD ruled short--distance behaviour of the vector and axial form 
factors in the large--$N_C$ limit (approximated with only one octet
of vector resonances) constrains the couplings of 
${\cal L}_{\rm R\chi T}$ in Eq.~(\ref{eq:ret}), which must
satisfy \cite{bi:11}~:
\begin{eqnarray}
\label{fvgv}
1 \, - \, \Frac{F_V \, G_V}{F^2} & = & 0 \; \; , \nonumber \\
2 F_V G_V \, - \, F_V^2 \, & = & 0 \; \; .
\end{eqnarray} 
In addition, the first Weinberg sum rule, in the limit 
where only the lowest narrow resonances contribute to the vector 
and axial--vector spectral functions, leads to
\begin{equation}
\label{fvfa}
F_V^2 - F_A^2 = F^2 \;.
\end{equation}
In this way all three couplings $F_V$, $G_V$ and $F_A$
can be written in terms of the pion decay constant~: 
$F_V = \sqrt{2} F$, $G_V = F/\sqrt{2}$ and $F_A = F$. These results are
well satisfied phenomenologically and we have adopted them \footnote{A
more thorough study of this procedure is given in Ref.~\cite{nos}.
}.
\begin{figure*}[!th]
\hspace*{0.5cm}
\includegraphics*[angle=-90,scale=0.65,clip]{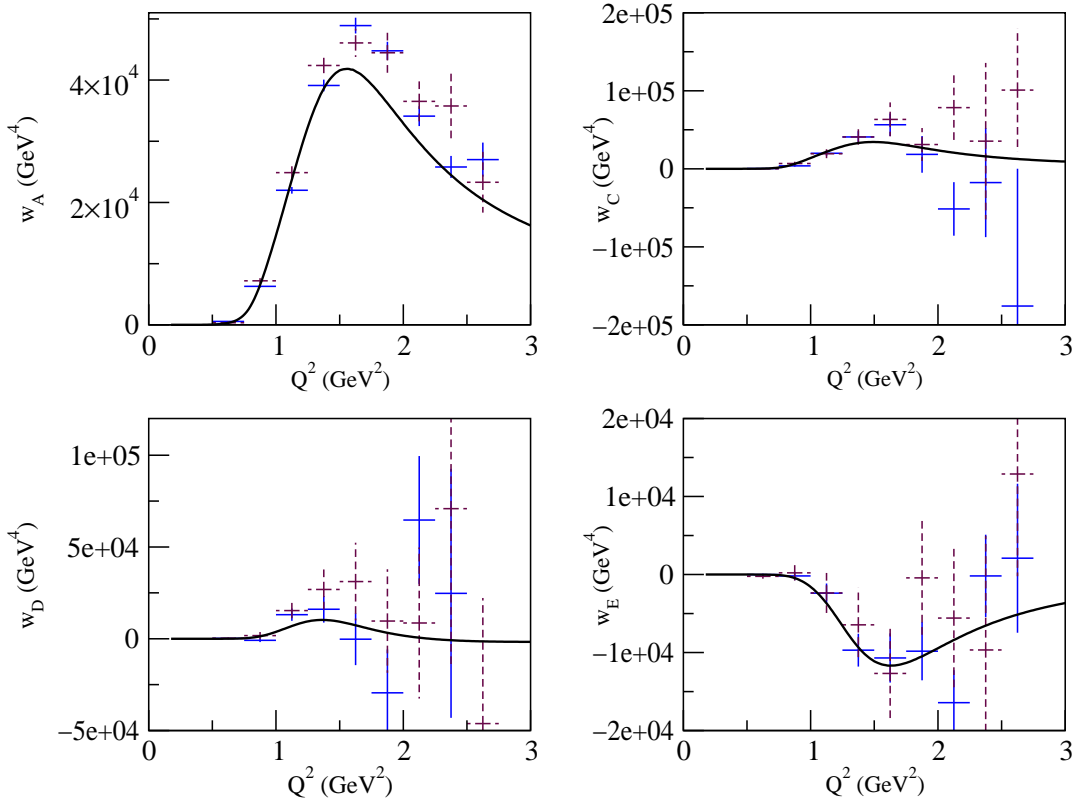}
\vspace*{-0.8cm}
\caption{\label{fig:f4} Theoretical values for the $w_A$, $w_C$, $w_D$
and $w_E$ integrated structure functions in comparison with the experimental
data from CLEO-II (solid) and OPAL (dashed) \protect{\cite{bi:8}}.}
\vspace*{-0.5cm}
\end{figure*}
The proper QCD driven behaviour of the $T_{\pm \mu}$ form factor imposes
in addition the constraints~:
\begin{eqnarray}
2 \, \lambda' \, - \, 1 & = & 0 \; \;, \nonumber \\
\lambda'' & = & 0 \; \;.
\label{rellam}
\end{eqnarray}

\section{Phenomenology of $\tppp$ processes}

To analyse the experimental data we will only consider the dominating
$J^P = 1^+$ driven axial--vector form factors, that satisfy
$T_{+\mu} = - T_{-\mu}$ hence providing the same predictions for both
$\tpppc$ and $\tpppn$ processes in the isospin limit.
\par
We have fitted the experimental values for the $\tpppc$ branching ratio
and normalized spectral function obtained by ALEPH \cite{bi:6} and 
we get a reasonable 
$\chi^2/d.o.f. = 64.5 / 52$ shown in Fig.~\ref{fig:f3}. 
Hence we get the axial--vector
$a_1(1260)$ parameters $M_A = (1.204 \pm 0.007) \, \mbox{GeV}$ and 
$\Gamma_{a_1}(M_A^2) = (0.48 \pm 0.02) \, \mbox{GeV}$, where the errors
are only statistical. We predict, accordingly, the structure functions 
\cite{bi:14} that we compare with the experimental results for 
$\tpppn$ in Fig.~\ref{fig:f4}.

\noindent {\bf Acknowledgements}  \\
We wish to thank S.~Narison and his team for the organization
of the QCD03 Conference. This work has been supported in part by 
TMR EURIDICE, EC Contract No. 
HPRN-CT-2002-00311, by MCYT (Spain) under grant
FPA2001-3031, by Generalitat Valenciana under grant GRUPOS03/013, by
the Agencia Espa\~nola de Cooperaci\'on Internacional (AECI) 
and by ERDF funds from the EU Commission.

\end{document}